\documentclass[a4paper,11pt]{article}
\usepackage{pos}

\usepackage{graphicx}
\usepackage{siunitx}
\usepackage{float}
\usepackage{booktabs}
\usepackage{tikz}
\usetikzlibrary{arrows, arrows.meta, 3d, calc}
\usepackage{pgfplots}
\pgfplotsset{compat=1.17}

\newcommand{\beeq}{\begin{equation}}
	\newcommand{\eneq}{\end{equation}}
\let\[\beeq
\let\]\eneq

\newcommand{\amu}{a_\mu}
\newcommand{\oa}{\omega_a}

\newcommand{\opp}{\omega'_p}
\newcommand{\oppt}{\tilde\opp}

\title{Magnetic Field Analysis for Fermilab Muon $g-2$}

\manuallySeparateAuthors
\author*[a]{A. Tewsley-Booth}
\author{\emph{for the Muon $g-2$ collaboration}}
\affiliation[a]{University of Kentucky}
\emailAdd{atewsleybooth@uky.edu}

\abstract{
The Muon g-2 experiment E989 at Fermilab measures the anomalous magnetic moment of the muon $\amu$ with improved precision compared to the Brookhaven (E821) experiments. The Brookhaven results are in tension with the Standard Model by more than $3\sigma$. The determination of $\amu$ requires the measurement of both the muon anomaly frequency, $\oa$, and the magnetic field, $\mathbf{B}$, that confines muons in a storage ring. The field is monitored by a set of coordinated nuclear magnetic resonance (NMR) measurements. NMR probes at fixed locations above and below the storage region constantly monitor the field. An in-vacuum trolley equipped with 17 NMR probes maps the muon storage region, and a special water-based NMR probe provides the calibration for the trolley probes. This presentation focuses on the determination of the time-dependent field maps from combining the fixed probe measurements and the trolley maps. The field maps are combined with the muon distribution to derive the average field observed by the muons during the measurement. These proceedings will cover the analysis from the first data run.
}

\FullConference{%
  *** The 22nd International Workshop on Neutrinos from Accelerators (NuFact2021) ***\\
  *** 6–11 Sep 2021 ***\\
  *** Cagliari, Italy ***}

\begin{document}
\maketitle

\section{Introduction}

Our collaboration reported a new measurement of the muon magnetic anomaly, $a_\mu = \frac{1}{2}(g_\mu - 2)$, based on our Run-1 data set, which was collected between March and July of 2018. The result is \[a_\mu(\mathrm{FNAL}) = 116592040(54) \times 10^{-11},\] measured to a precision of 460 ppb, and is in good agreement with the previous measurement from Brookhaven \cite{Bennett2006}. A set of four companion papers cover the final result \cite{gm2prl2021}, the anomalous spin precession frequency \cite{gm2omegaa2021}, the magnetic field measurements \cite{gm2field2021}, and the beam dynamics corrections \cite{gm2bd2021}. These proceedings will broadly cover the magnetic field measurement and analysis for the Run-1 data set.\footnote{A similar talk was given by the speaker at the 2021 DPyC-SMF conference.}

The spin of particle with a magnetic moment precesses about an external non-parallel magnetic field at the Larmor frequency \cite{Jackson}, \[\omega = \frac{gq}{2m} B\] . If the particle's charge, $q$, and its mass, $m$, are known, simultaneous measurement of the spin precession frequency $\omega$ and the magnetic field $B$ would allow the determination of the g-factor of the particle. The muon's magnetic moment is slightly greater than two due to interactions with the vacuum. This shift, the magnetic anomaly $a = \frac{g-2}{2}$, is effected by interactions with virtual particles in the vacuum. Therefore, a measurement of the magnetic anomaly can be used to probe properties of these interactions. Current $g-2$ theory takes into account QED, electroweak, and QCD. The QED terms dominate the value, but the QCD terms dominate the uncertainty \cite{gm2whitepaper2020}.

In a constant magnetic field, the relative angle between the momentum and spin vectors of the muon evolves over time. The rate of change of this angle, called the anomalous precession frequency $\oa$, is (nearly) proportional to the magnetic anomaly, $\amu$. \[\vec\oa = \vec\omega_s - \vec\omega_c \approx -\frac{q}{m} \amu \vec B, \label{eq:oa-simple}\] where $\omega_s$ is the spin precession frequency and $\omega_c$ is the cyclotron (momentum precession) frequency. Simultaneous measurement of $\oa$ and the magnetic field enable determination of $\amu$. In this experiment, the field is measured as $\oppt$, the Larmor frequency of the proton in water in the same magnetic field, weighted by the muon distribution.

The muon-averaged field, $\oppt$, can be laid out schematically as \[\oppt = f_\text{field} ~\left\langle \omega_\mathrm{p, meas} \bigotimes \rho_\mu \right\rangle ~(1 + B_{q} + B_{k}). \label{eq:schematical}\] Here, $f_\mathrm{field}$ is the absolute calibration of the field that takes $\omega_\text{p, meas}$ to $\opp$, the equivalent Larmor frequency of the proton in water. The term $\left\langle \omega_\mathrm{p, meas} \bigotimes \rho_\mu \right\rangle$ is the average measured field weighted by the muon distribution $\rho_\mu$. The terms $B_q$ and $B_k$ are the effects of fast transient magnetic fields that cannot be measured by the field systems that measure $\omega_\text{p, meas}$.

\section{Measuring \texorpdfstring{$\omega_p$}{omega\_p}}
\label{sec:field}

The magnetic field experienced by the muons is measured using several different magnetometer systems. The most important of these are the calibration probe, trolley, and fixed probe systems. The primary field mapping system, the trolley, is an array of magnetometers located in the vacuum chamber. The trolley is pulled by cable around the muon storage region, measuring the field as it goes. Each of these systems uses NMR magnetometry, and together they form a calibration chain that allows us to measure the absolute field to a precision of 114 ppb (56 ppb from the calibration, measurements, analysis, and averaging; and 99 ppb from the effects of fast transient fields).

\begin{figure}[htb]
	\centering
	\includegraphics[width=0.45\linewidth]{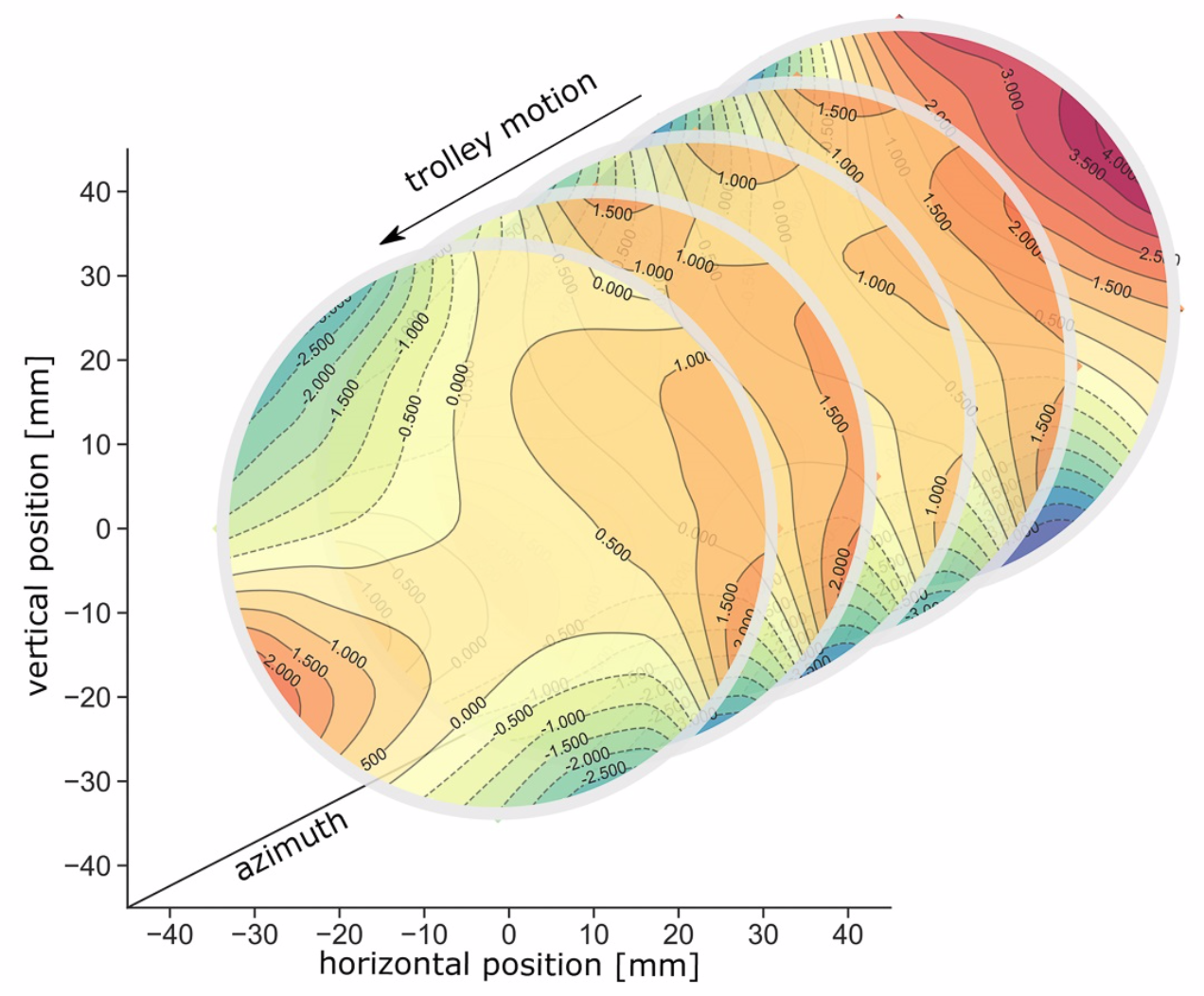}
	\caption{The trolley generates precise field maps as a function of azimuthal position as it travels around the ring. Between trolley runs, these maps are interpolated using measurements from the fixed probes that track the evolution of the field.}
	\label{fig:trolley-slices}
\end{figure}

The calibration chain begins at Argonne National Laboratory in a precision MRI magnet. There, the calibration probe is cross-checked with a He-3 NMR probe \cite{Farooq2020}. The two probes were found to be in excellent agreement. The calibration probe is then transferred to the storage ring at FNAL, where the calibration from the calibration probe is transferred to each of the trolley probes. Then, as the trolley moves about the storage ring, it can transfer the calibration to the array of fixed probes.

\begin{figure}[htb]
	\centering
	\includegraphics[width=0.3\linewidth]{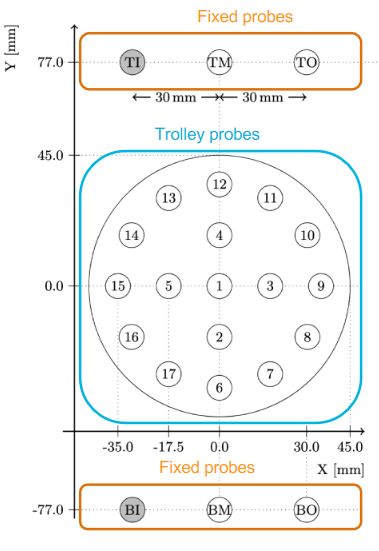}
	\caption{The relative positions of the fixed and trolley probes in an azimuthal slice of the ring. In this coordinate system, the muon's magic radius is at $(0,0)$.}
	\label{fig:probe-locations}
\end{figure}

After calibration, the primary systems used to map the magnetic field in the storage ring are the trolley and the fixed probes. The trolley makes its measurements in the same volume that the muons fill. It has 17 NMR probes and takes measurements in about 4000 azimuthal locations around the ring, as shown in Figure \ref{fig:trolley-slices}. However, the trolley cannot be operated during muon fills because it blocks their path, so it is pulled out of the way into its garage most of the time, only mapping the field about every three days. On the contrary, the fixed probes only measure at 72 azimuthal locations and are physically located outside the vacuum chambers, but they can continue measuring the field during muon fills, providing information about how the field evolves between trolley runs. Figure \ref{fig:probe-locations} shows the relative locations of the trolley and fixed probes in an azimuthal slice of the ring. The two sources are combined in the analysis, with the fixed probe data being used to interpolate the field map between the trolley runs. For Run-1, this procedure, from measurement through analysis, accounts for 56 ppb of the total uncertainty.

The NMR magnetometers are good for measuring quasi-static fields, but transients with characteristic times faster than about a second require special systematic studies to measure. There are two primary sources of transient magnetic fields in the storage ring: the electrostatic quadrupoles and the faster kickers. The ESQ plates are charged to high voltage at each muon injection to provide vertical focusing on the beam. Charging the plates induces vibrations, which generate a magnetic field that perturbs the muons. For the Run-1 analysis, the effect of this transient field was found to be -17 ppb with an uncertainty of 92 ppb, making it the dominant systematic effect on the final result. Since the Run-1 publication, we have completed a more in-depth systematic study and expect this uncertainty to be reduced in future results. Furthermore, beginning in Run-5, we have changed the ESQ charging procedure to reduce the induced vibrations that cause the transient.

\begin{figure}[htb]
	\centering
	\begin{tikzpicture}[scale=0.75]
		\node at (0,-0.25) {\includegraphics[width=7.5cm]{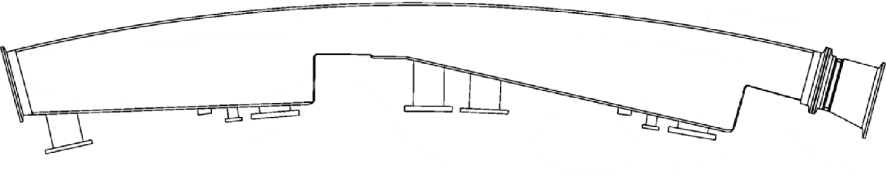}};
		\draw [very thick, dotted, color=blue] (-4.9, -0.1) arc (103:74.75:20);
\draw [blue] (-3.66, 0.15) -- ++(0,-1.5) node [anchor=west, blue] {\tiny Muon Orbit};

\draw [very thick, fill] (0,0) circle (0.025);


\draw [ultra thick, cyan] (-13:0.875) -- ++(75:0.375);
\draw [ultra thick, cyan] (-13:0.8375+0.3141) -- ++(75:0.4);
\draw [ultra thick, cyan] (-13:0.8375+2*0.3141) -- ++(75:0.425);
\draw [ultra thick, cyan] (-13:0.8375+3*0.3141) -- ++(75:0.45);
\draw [ultra thick, cyan] (-13:0.8375+4*0.3141) -- ++(75:0.475);
\draw [ultra thick, cyan] (-13:0.8375+5*0.3141) -- ++(75:0.5);
\draw [ultra thick, cyan] (-13:0.8375+6*0.3141) -- ++(75:0.525);
\draw [ultra thick, cyan] (-13:0.8375+7*0.3141) -- ++ (75:0.55);
\draw [cyan] (-13:0.8375+3*0.3141) -- ++(0, -1) node [anchor=west, cyan] {\tiny Straw Trackers};

\draw [thick, cyan, fill=cyan!30!white] (-13:3.42) -- ++(75:0.5) -- ++(-15:0.5) -- ++(255:0.5) -- (-13:3.42);
\draw [cyan] (3.8,-0.88) -- ++(0,-0.5) node [anchor=west, cyan] {\tiny Calorimeter};

\draw [Rays-{Rays[n=7]}, red, thick, dashed] (-3,0.25) arc (92:73:21);
\draw [red] (-2.92,0.25) -- ++(0,1.25) node [anchor=west, red] {\tiny High Momentum Decay Positron};

\draw [Rays-{Rays[n=7]}, red, thick, dashed] (0,0.4) arc (85:60:9);
\draw [red] (0.1,0.4) -- ++(0,0.75) node [anchor=west, red] {\tiny Low Momentum Decay Positron};
	
	\end{tikzpicture}
	\caption{Decay positrons travel through the straw trackers. Their paths can be traced back using knowledge of the magnetic field to their decay vertices, allowing for a calculation of the muon distribution.}
	\label{fig:tracker-cartoon}
\end{figure}
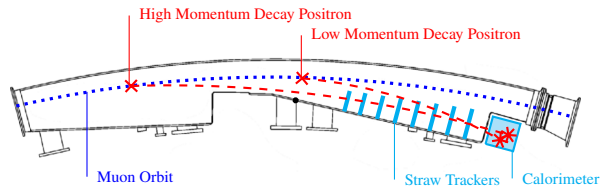

The source of the other transient is the kicker system, which uses a fast magnetic field at the beginning of each muon injection to kick the muons onto their ideal orbit. The kick induces eddy currents that perturb the field as they decay. This effect is measured using a Faraday magnetometer that can measure the field at the nanosecond level. The average effect on the muons was found to be -27 ppb, with an uncertainty of 37 ppb. This measurement has also been refined and repeated since the Run-1 publication to reduce its associated uncertainty. Table \ref{tab:uncertainty} details the corrections and uncertainties for the quasi-static and transient dynamics of the field, weighted by the muon distribution.

In order to calculate the average magnetic field experienced by the muons, we need to know both the magnetic field and the muon distribution in the storage ring as a function of position and time. The muon distribution is measured by the straw trackers at two azimuthal locations. Those two distributions are then used to extrapolate the distribution around the whole ring by combining the measurement with beam dynamics simulations.

The straw trackers are formed of layers of overlapping straws filled with gas that is ionized as positrons travel through the device towards the calorimeters. By measuring the positions where the ionization occurs in the straws, we can find the tracks the positrons took from their decay positions (see Figure \ref{fig:tracker-cartoon}). Extrapolating backwards through the magnetic field, these tracks are used to determine the decay vertices of the positrons, which are used as a proxy for the muon distribution at the two azimuthal locations of the straw trackers. We use the distributions averaged over several hours of data collection as a weighting function when we average the magnetic field's non-uniformity.

\begin{table}[htb]
	\centering
	\begin{tabular}{lrr}
		\midrule\midrule
		Quantity & Correction Terms & Uncertainty \\
		& (ppb) & (ppb) \\
		\midrule
		$\oppt$ & -- & 56 \\
		$B_k$ & -27 & 37 \\
		$B_q$ & -17 & 92 \\
		\midrule
		Totals & -44 & 114\\
		\midrule\midrule
	\end{tabular}
	\caption{The corrections and uncertainties to the field terms in Equation \ref{eq:schematical}.}
	\label{tab:uncertainty}
\end{table}

\section{Acknowledgments}

We thank the Fermilab management and staff for their strong support of this experiment, as well as the tremendous support from our university and national laboratory engineers, technicians, and workshops. The Muon $g-2$ Experiment was performed at the Fermi National Accelerator Laboratory, a U.S. Department of Energy, Office of Science, HEP User Facility. Fermilab is managed by Fermi Research Alliance, LLC (FRA), acting under Contract No. DE-AC02-07CH11359. Additional support for the experiment was provided by the Department of Energy offices of HEP and NP (USA), the National Science Foundation (USA), the Istituto Nazionale di Fisica Nucleare (Italy), the Science and Technology Facilities Council (UK), the Royal Society (UK), the European Union's Horizon 2020 research and innovation programme under the Marie Sk\l{}odowska-Curie grant agreements No. 690835, No. 734303, the National Natural Science Foundation of China (Grant No. 11975153, 12075151), MSIP, NRF and IBS-R017-D1 (Republic of Korea), the German Research Foundation (DFG) through the Cluster of Excellence PRISMA+ (EXC 2118/1, Project ID 39083149). 

\pagebreak

\end{document}